\documentclass[aps,prl,amsmath,amssymb,twocolumn, reprint]{revtex4-1}

\usepackage{graphicx}
\usepackage{dcolumn}
\usepackage{footmisc}
\usepackage[normalem]{ulem}
\usepackage{bm}
\usepackage{hyperref}
\usepackage{xcolor}
\usepackage{float}

\begin{document}
\setlength{\abovedisplayskip}{5pt}
\setlength{\belowdisplayskip}{5pt}
\setlength{\abovedisplayshortskip}{5pt}
\setlength{\belowdisplayshortskip}{5pt}
\hyphenpenalty=1050
\hyphenation{KATRIN}

\newcommand{\gh}[1]{\textcolor{blue}{[{\bf GH}: #1]}} 
\newcommand{\mc }[1]{\textcolor{orange}{#1}}
\newcommand{\com}{\textcolor{red}}

\preprint{}

\title{Upper Limits on the Cosmic Neutrino Background from Cosmic Rays}

\author{Mar Císcar-Monsalvatje\textsuperscript{1}, Gonzalo Herrera\textsuperscript{2,1,3}, Ian M. Shoemaker\textsuperscript{2}}
\affiliation{
\textsuperscript{1}School of Natural Sciences, Technische Universit\"at M\"unchen, James-Franck-Stra\ss{}e, 85748 Garching, Germany,\\ \textsuperscript{2} Center for Neutrino Physics, Department of Physics, Virginia Tech, Blacksburg, VA 24061, USA,\\ \textsuperscript{3} Max-Planck-Institut f\"ur Physik (Werner-Heisenberg-Institut), F\"ohringer Ring 6,80805 M\"unchen, Germany,}

\begin{abstract}
Extragalactic and galactic cosmic rays scatter with the cosmic neutrino background during propagation to Earth, yielding a flux of relic neutrinos boosted to larger energies. If an overdensity of relic neutrinos is present in galaxies, and neutrinos are massive enough, this flux might be detectable by high-energy neutrino experiments. For a lightest neutrino of mass $m_{\nu} \sim 0.1$ eV, we find an upper limit on the local relic neutrino overdensity of $\sim 10^{13}$ and an upper limit on the relic neutrino overdensity at TXS 0506+056 of $\sim 10^{10}$. Future experiments like GRAND or IceCube-Gen2 could improve these bounds by orders of magnitude.
\end{abstract}

\maketitle

\section{Introduction}

The cosmic neutrino background is present in our Universe with a total average density of $n_{\nu} \sim 336$ cm$^{-3}$ for all flavors \cite{Dolgov:1997mb, Mangano:2005cc}. Despite their large density, their energies are lower than the largest neutrino mass \cite{Vitagliano:2019yzm}, which is constrained by the KATRIN experiment for electron antineutrinos to $m_{\nu} < 0.8$ eV \cite{KATRIN:2021uub}. This makes their detection very challenging. Relic neutrinos can hardly induce detectable signals at Earth-based detectors since their interaction cross section at sub-eV energies is suppressed, and the energy depositions are very small, $\sim$ meV$-$eV. A widely discussed mechanism to overcome the low energy deposition is the capture of a neutrino on a $\beta-$unstable nucleus, which is a thresholdless reaction \cite{PhysRev.128.1457,Cocco:2007za,Blennow:2008fh,Long:2014zva,PTOLEMY:2018jst,Akita:2020jbo,Alvey:2021xmq,Bauer:2022lri, Perez-Gonzalez:2023llw}. However, the feasibility of this detection method is limited by Heisenberg's uncertainty principle \cite{Cheipesh:2021fmg,Nussinov:2021zrj,PTOLEMY:2022ldz}. Given the difficulties of such detection, some works in the literature have invoked Beyond the Standard Model (BSM) mechanisms enabling relic neutrinos to be detectable at Earth by various mechanisms and with different signatures and experiments, \textit{e.g.} \cite{Weiler:1982qy,Fargion:1997ft,Ibe:2014pja,Brdar:2022kpu, Das:2022xsz, Asteriadis:2022zmo, Yin:2017wxm}.

In this letter, we explore the possibility that cosmic rays scatter off relic neutrinos and boost them via purely Standard Model mechanisms, inducing a flux of relic neutrinos with larger energies reaching the Earth. A similar task was done in the 80s when the neutrino mass was less constrained, and the uncertainties in the atmospheric and astrophysical high-energy neutrino fluxes were larger \cite{1980PThPh..64.1089H, Hara:1980mz}. Here, we estimate this flux due to scatterings of cosmic ray protons with massive neutrinos and assess whether such a contribution could be discerned from other neutrino sources. In particular, we will calculate the flux of relic neutrinos boosted by cosmic ray protons in the Milky Way and the flux of relic neutrinos boosted by cosmic ray protons in the blazar TXS 0506+056, which is believed to be a proton accelerator from its detected emission in high-energy neutrinos \cite{IceCube:2018cha}. We will place an upper limit on the local, or TXS 0506+056 present relic neutrino overdensity from a non-observation of a particular spectral feature on top of the measured atmospheric and astrophysical high energy neutrino fluxes, or from an excess of events w.r.t experimental limits on cosmogenic neutrinos. We further analyze the dependence of this constraint on the neutrino mass and discuss whether such overdensities could be obtained in a model where neutrinos self-interact via a light scalar mediator \cite{Smirnov:2022sfo}. The phenomenological probe considered in this work allows the testing of neutrino overdensities on much larger physical scales than other complementary probes such as \cite{KATRIN:2022kkv, Tsai:2022jnv}.

\section{Boosted flux of relic neutrinos from cosmic ray proton scatterings}

Cosmic ray protons have much larger energies than the relic neutrinos, which are effectively at rest. The maximum kinetic energy transferred to a neutrino in a single collision is \cite{Bringmann_2019, Cappiello:2018hsu}

\begin{equation}\label{eq:Tmax}
T_\nu^{\max }=\frac{T_p^2+2 m_p T_p}{T_p+\left(m_p+m_\nu\right)^2 /\left(2 m_\nu\right)},
\end{equation}
where $T_p$ is the kinetic energy of the proton. The induced flux of cosmic relic neutrinos by scatterings with cosmic ray protons 
is given by the following line of sight integral

\begin{equation}
\frac{d \Phi_\nu}{d T_p}=\int \frac{d \Omega}{4 \pi} \int_{\text {l.o.s.}} d \ell \sigma_{p \nu}(T_{p}) n_{\nu} \frac{d \Phi_p}{d T_p} \equiv \sigma_{\nu p}(T_p) n_{\nu} \frac{d \Phi_p}{d T_p} D_{\mathrm{eff}},
\end{equation}
where $\sigma_{\nu p}(T_p)$ refers to the neutral current proton-neutrino scattering cross section, $d \Phi_p/d T_p$ is the proton flux spectrum, and $D_{\rm eff}$ is the effective distance. For the Milky Way case we consider the measured cosmic ray proton flux on Earth up to the GZK cut-off ($T_p\sim5\times 10^{19}$~eV) \cite{Greisen, Zatsepin, Cronin:1997vy}. We assume $D_{\rm eff} \sim 10$ kpc over which cosmic rays scatter off relic neutrinos in the Milky Way. In reality, ultra-high energy cosmic rays could scatter off relic neutrinos over larger distances, of the order of $10-100$ Mpc \cite{Plotko:2022urd}, so our choice is conservative. For TXS 0506+056, we calculate the high-energy proton flux following \cite{Wang:2021jic}, with a cut-off at $1.7\times 10^{9}$ GeV \cite{MAGIC:2018sak, Winter:2019hee}, and rescale it by $1/d^{2}$, where $d=1835.4$ Mpc. We assume that protons scatter off relic neutrinos over an effective distance of $D_{\rm eff}=10$ kpc in the host galaxy of TXS 0506+056, which lies under the mean free path of these protons for the relic neutrino overdensities considered in this work.

\begin{figure}[H]
    \centering
\includegraphics[trim={0 0 0cm 0},clip, width=1\linewidth]{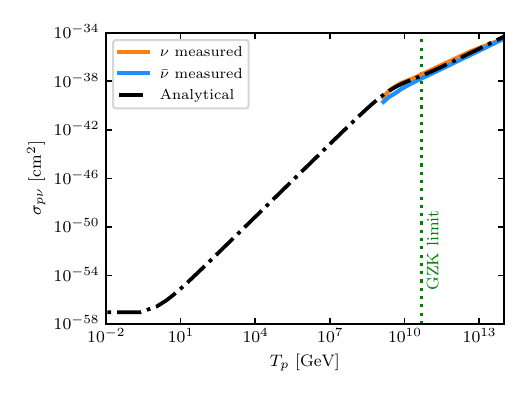}
    \caption{Average neutrino/antineutrino-proton scattering cross section used in this work (dash-dotted black line) compared with the best-fit experimentally measured values of the total scattering cross section at high energies. We recast the cross section measurements shown in \cite{Formaggio:2012cpf} in the frame where the neutrino is at rest, for a neutrino mass of $m_{\nu}=0.1$ eV. Furthermore, we show the maximum proton kinetic energy considered in this work, corresponding to the GZK limit.}
\label{fig:cross_section}
\end{figure}

The neutral current neutrino-proton cross section in the regime of interest where the momentum transfer is smaller than the mediator mass scales with energy as 
\begin{equation}
    \sigma_{p\nu}\sim\begin{cases}
        \frac{G_{F}^{2} s}{\pi} & \mathrm{for}\,\,\, 2m_\nu E_p> m_p^2,\\
         \frac{G_{F}^{2} E_p^2 m_\nu^2}{\pi m_p^2} &\mathrm{for}\,\,\, 2m_\nu E_p< m_p^2,
    \end{cases}
\end{equation}
where $s=2m_{\nu}E_{p} + m_p^2 + m_\nu^2$ and $E_p = T_p + m_p$. The dependence at very high $s$ is not relevant in this work due to the high suppression by the small value of $m_\nu$. We have checked that our expressions for the cross section match the experimentally measured values at sufficiently high center of mass energies \cite{Formaggio:2012cpf}, see Fig.~\ref{fig:cross_section}.

The cosmic neutrino flux as a function of the final neutrino energy reads
\begin{equation}
\frac{d \Phi_\nu}{d T_\nu}=\int_0^{\infty} d T_p \frac{d \Phi_\nu}{d T_p} \frac{1}{T_\nu^{\max }\left(T_p\right)} \Theta\left[T_\nu^{\max }\left(T_p\right)-T_\nu\right].
\end{equation}

\begin{figure}[t]
    \centering
\includegraphics[width=\linewidth]{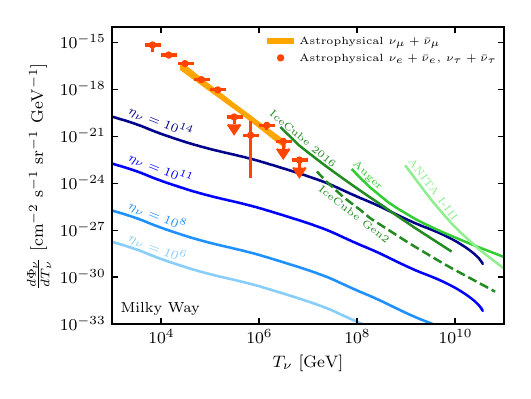}
\includegraphics[trim={0 0 0.2cm 0},clip, width=1\linewidth]{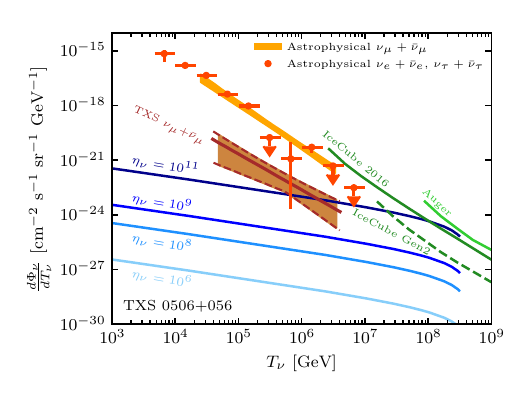}
    \caption{\textit{Upper plot}: Flavor averaged ($\nu_\alpha+\bar\nu_\alpha$) flux of relic neutrinos boosted by Milky Way cosmic ray protons at Earth, for different values of the local neutrino overdensity in the Milky Way. The neutrino mass is taken to be $m_\nu=0.1$ eV. For comparison, we show the flux from the sum of atmospheric neutrinos and diffuse extragalactic neutrinos measured by experiments at Earth \cite{Super-Kamiokande:2015qek, IceCube:2016umi, IceCube:2020acn}. We also show the current upper limits on the cosmogenic neutrino flux at very high-energies from Auger \cite{PierreAuger:2015ihf}, IceCube \cite{IceCube:2016uab} and ANITA \cite{ANITA:2018vwl}. We also show sensitivity projections from GRAND200k (10yr) and IceCube-Gen2 radio (10 yr) \cite{Kotera:2021hbp, IceCube-Gen2:2021rkf}. \textit{Lower plot}: Flavor averaged ($\nu_\alpha+\bar\nu_\alpha$) boosted relic neutrino flux from TXS 0506+056 \cite{IceCube:2018cha}. We also show in brown the observed non-diffuse flux of neutrinos coming from TXS 0506+056 (see main text for details).}
\label{fig:boosted_relic_nus}
\end{figure}

Using this formalism, we compute the fluxes of relic neutrinos boosted by cosmic ray protons from the Milky Way and TXS 0506+056. The boosted flux is orders of magnitude smaller than the flux of atmospheric and high-energy neutrinos. However, this flux might become sizable at energies above $10^5-10^6$ GeV for sufficiently large neutrino overdensities, $\eta_\nu$, in the Milky Way or TXS 0506+056, as can be seen in Fig.~\ref{fig:boosted_relic_nus}. 

While in standard scenarios, the neutrino clustering via gravitational effects is expected to be very small \cite{Ringwald:2004np, Mertsch:2019qjv, Zimmer:2023jbb, Hotinli:2023scz}, in some models, neutrinos can self-interact via a scalar mediator, which leads to large overdensities. The maximal predicted neutrino density in this scenario is $n_{\nu} \sim 4.3 \times 10^{8}$ cm$^{-3}$~\cite{Stephenson:1996qj,Smirnov:2022sfo} for $m_{\nu} \sim 0.1$ eV, which translates into an overdensity of $\eta_\nu \sim 1.2 \times 10^6$. Currently, the most stringent upper limit on the local neutrino overdensity w.r.t. the average value comes from the KATRIN experiment, which is $\eta_\nu \leq 9.7 \times 10^{10}$ at 90$\%$ CL \cite{KATRIN:2022kkv}. 

It can be noticed from the upper plot of Fig.~\ref{fig:boosted_relic_nus} that for overdensities at the level of the KATRIN bound, the boosted relic neutrino flux in the Milky Way does not reach the measurements reported by the IceCube collaboration on the diffuse extragalactic neutrino background (shown in light orange for muon neutrinos and red points for electron and tau neutrinos). However, for TXS 0506+056, such an overdensity would induce a detectable neutrino flux above IceCube measurements in the source direction, and also above the non-observation limits placed on cosmogenic neutrinos (green solid lines). As discussed, we do not speculate how such large relic neutrino densities would be present at the source. Similar discussions have been placed in the context of dark matter spikes around the supermassive black hole of TXS 0506+056, with enhancements above $10^9$ times the cosmological density of dark matter \cite{Gondolo:1999ef, Wang:2021jic, Ferrer:2022kei, Cline:2022qld}. However, to the best of our knowledge, a discussion on the clustering of relic neutrinos around supermassive black holes is lacking in the literature.

As shown in Fig.~\ref{fig:boosted_relic_nus}, the relic neutrinos may contribute at the energies relevant for the diffuse extragalactic and cosmogenic neutrino fluxes but with a different spectral index. Thus, if a sufficiently large overdensity is present, the IceCube experiment could observe the relic neutrino contribution by analyzing the spectral features.

It is worth mentioning that here we have focused on the boosted relic neutrino contribution arising from cosmic ray proton scatterings, but relic neutrinos are also boosted to larger energies by scatterings off cosmic ray electrons. Due to their lower density and energies, however, we have found this contribution to be negligible compared to that from cosmic ray protons.

\section{Upper limit on local relic neutrino overdensity}

We can set an upper limit on the relic neutrino overdensity from the requirement that the boosted relic neutrino flux should not be larger than the measured diffuse extragalactic neutrino flux, or the existing limits for cosmogenic neutrinos. In particular, we follow the criteria that the sum of boosted relic neutrino flux shall not be larger than the observed neutrino flux at any energy by a factor $C$
\begin{equation}
\Phi_{\rm C \nu B}^{\rm Boosted} \leq C \Phi_{\nu}^{\rm Observed},
\end{equation}
where the factor $C$ depends on the uncertainty of the measured neutrino flux, which is energy dependent. Uncertainties in the astrophysical neutrino fluxes are below one order of magnitude at all relevant energies, except for one energy bin of IceCube corresponding to electron and tau neutrinos around $10^6$ GeV \cite{IceCube:2022der}. Thus, we will consider $C=1$.

The maximal neutrino density in a cluster due to Yukawa interactions induced by a massless scalar mediator is determined by the neutrino mass as \cite{Smirnov:2022sfo}
\begin{equation}
n_\nu^{\max }=4.3 \times 10^{8} \mathrm{~cm}^{-3}\bigg(\frac{m_\nu}{0.1\, \mathrm{eV}}\bigg)^3.
\end{equation}
This corresponds to an overdensity, w.r.t. to the standard average value of
\begin{equation}\label{eq:nuclusteringlimit}
\eta^{\max }=1.2 \times 10^6 \bigg(\frac{m_\nu}{0.1\, \mathrm{eV}}\bigg)^3.
\end{equation}

\begin{figure}[t!]
    \centering
\includegraphics[width=\linewidth]{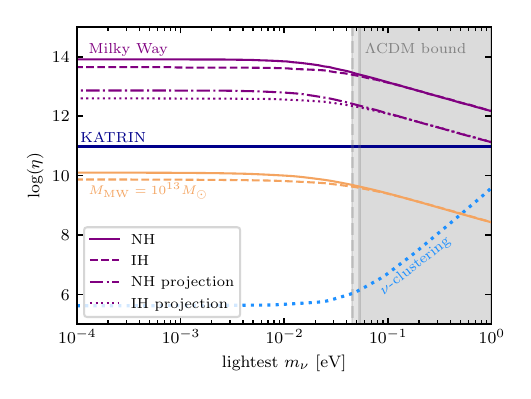}
\includegraphics[width=\linewidth]{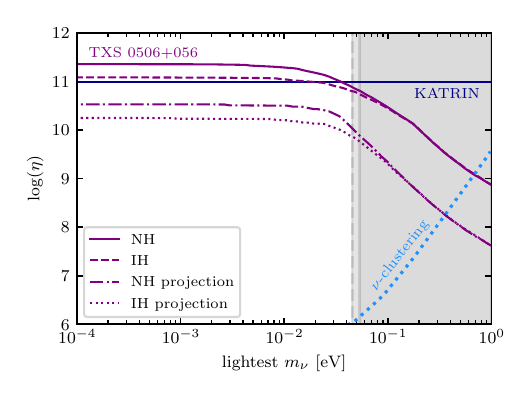}
    \caption{\textit{Upper plot}: Constrained local overdensity as a function of the lightest neutrino mass, for an effective distance of $D_{\rm eff}=10$ kpc and for both normal hierarchy (solid line) and inverted hierarchy (dashed line). We also show projected constraints from future experiments in dashed-dotted and dotted lines. For comparison, we show the bound from KATRIN and maximum values favored by relic neutrino clustering mediated by self-interactions. We further show the cosmological bound on the sum of neutrino masses from the combination of CMB and BAO data at 68$\%$ C.L, ($\sum m_{\nu} < 0.19$ eV) for NH (solid) and IH (dashed). In orange, we show the requirement for the neutrino halo mass to not exceed $M_{\rm MW} \lesssim 10^{13} M_{\odot}$. \textit{Lower plot}: Same for TXS 0506+056.}
\label{fig:overdensityvsmass}
\end{figure}
In Fig.~\ref{fig:overdensityvsmass}, we show our upper limit on the local neutrino overdensity (upper plot) and neutrino overdensity in TXS 0506+056 (lower plot).
We show this limit for a normal hierarchy of neutrino masses (NH) in solid purple and an inverted hierarchy (IH) in dashed purple as a function of the lightest neutrino mass. We also show in purple dash-dotted and dotted lines projections for the overdensities which future cosmogenic neutrino experiments IceCube-Gen2 and GRAND will be able to probe \cite{Kotera:2021hbp, IceCube-Gen2:2021rkf}.

For comparison, we show the previous bound by KATRIN in solid blue and the theoretically motivated values of the relic neutrino overdensities in the case of Yukawa self-interactions, following Eq.~(\ref{eq:nuclusteringlimit}), in dotted light blue, taking into account the contribution from the three neutrino eigenstates. 
The grey region (right of the grey solid line for NH and of grey dashed line for IH) is disfavoured in $\Lambda$CDM cosmology, \textit{e.g.} \cite{RoyChoudhury:2018gay,Chacko:2019nej,DiValentino:2021hoh, FrancoAbellan:2021hdb}.

Our limit from the Milky Way is weaker than the bounds from KATRIN and asteroid data \cite{KATRIN:2022kkv,Tsai:2022jnv}. However, these limits only apply at distance scales as long as 1 AU, while our bound applies up to a much larger scales. Furthermore, since the high-energy cosmic rays are likely of extragalactic origin, the total boosted relic neutrino flux on Earth could receive sizable contributions from scatterings in the intergalactic medium and therefore be significantly larger than estimated here. We leave this analysis for future investigation. 

Our limit from TXS 0506+056 is more stringent than the KATRIN bound for neutrino masses where the lightest neutrino mass is $m_{\nu} \gtrsim$ 0.04 eV, and the future projections improve it for any value of the neutrino masses. As expected, when the neutrino masses are higher, our constraint on the local relic overdensity becomes more stringent. This is due to the fact that it takes less energy transfer to boost the relic neutrinos to high energies, and the cosmic ray proton flux decreases at higher energy. This also explains why the IH bound is more stringent than the NH one, as in the IH case, the sum of neutrino masses will be higher for a determined lightest neutrino mass.

It should be stressed that for sufficiently large neutrino overdensities in the Milky Way extending over $D_{\rm eff}=10$ kpc, the mass of the neutrino halo could exceed the total mass of the Milky Way inferred from observations. Therefore, we impose an upper limit on the maximum neutrino overdensity in the Milky Way from the requirement that the neutrino halo mass must not exceed $M_{\rm MW} \lesssim 10^{13} M_{\odot}$, shown as an orange line in Fig.~\ref{fig:overdensityvsmass}. For TXS 0506+056, the total mass of the host galaxy is uncertain, but the limit would be close to the current one obtained from cosmic rays for $M_{\rm TXS} \lesssim 10^{14} M_{\odot}$. We emphasize, however, that the total mass of the neutrino halo will strongly depend on the radial dependence of the neutrino density profile, and the extension of the neutrino overdense region, which may not be equivalent to $D_{\rm eff}$, so the orange line in the figure should be taken with a grain of salt. These aspects are left for detailed investigation in future work.

We also note that for sufficiently small neutrino masses ($\sim10^{-4}$ eV), the mathematical treatment for the boosted flux would include modifications due to the finite C$\nu$B temperature \cite{Lunardini:2013iwa}. This is, however, not relevant for the derivation of our limits, as in the regime where the lightest neutrino is too light, the contribution of the heavier neutrinos dominates the limit. In the case of a reduced effective neutrino mass arising from matter effects in overdensities induced by neutrino self-interactions, our bounds would apply to the effective mass of neutrinos within the cluster.

\begin{figure}[t!]
    \centering
\includegraphics[width=\linewidth]{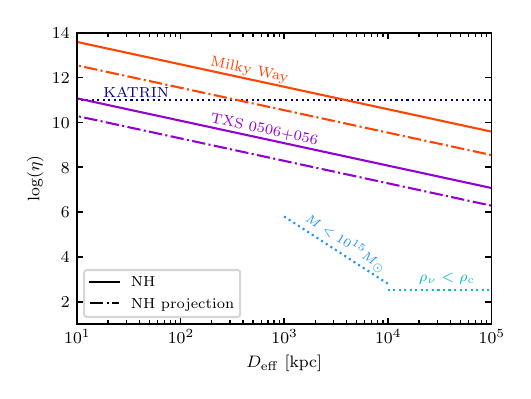}
    \caption{Constrained relic neutrino overdensity as a function of the effective distance $D_{\rm eff}$ travelled by cosmic rays measured at the Milky Way, and from the cosmic rays at TXS 0506+056. We plot them for a lightest neutrino mass of $m_\nu =0.03$ eV, lying at the limit of the cosmological constraint at 95$\%$ C.L, and for normal hierarchy (solid line) and projected constraints from future experiments also on normal hierarchy (dashed-dotted). For comparison, we show the bound from KATRIN. Furthermore, we show estimated bounds at large scales from the requirement that the contribution to the mass from the C$\nu$B should not exceed the typical mass of Galaxy clusters \cite{Bohringer:2017dro} (on scales from 1 to 10 Mpc) and that the neutrino background density does not exceed the critical density of the Universe (on scales larger than 10 Mpc).}
\label{fig:Deff}
\end{figure}

We have stressed that our constraints are linearly sensitive to the effective scale of the relic neutrino overdensity over which scatterings of cosmic rays occur. In Fig.~\ref{fig:Deff}, we show for comparison a limit on the relic neutrino overdensity as a function of the parameter $D_{\rm eff}$, for normal hierarchy and a lightest neutrino mass of $m_{\nu}=0.03$ eV, and confront it to the KATRIN bound, although in reality this limit only applies up to 2 AU $\sim 0.01$ pc scales. As can be seen in the figure, our limit on the relic neutrino overdensity from TXS 0506+056 can be as strong as $\eta \sim 10^{8}$ if ultra-high energy cosmic rays propagate over distances of 10 Mpc. Indeed, we have indications that ultra-high energy cosmic rays measured on Earth can propagate over distances of 10 to 100 Mpc.

It should also be noted that an indirect limit on the relic neutrino overdensity arises from Pauli exclusion principle, which disfavor relic neutrino overdensities larger than $\eta \gtrsim 10^{6}$ \cite{Bauer:2022lri, KATRIN:2022kkv, Bondarenko:2023ukx}.

\begin{figure}[t!]
    \centering
\includegraphics[width=\linewidth]{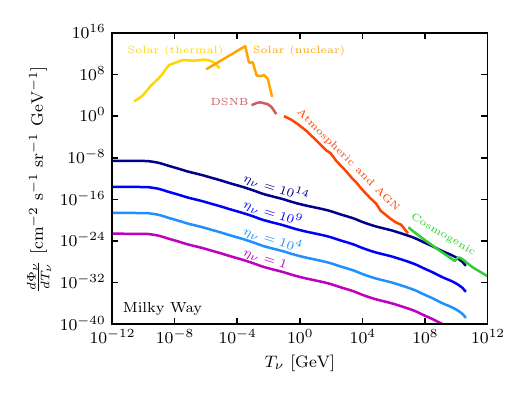}
\includegraphics[width=\linewidth]{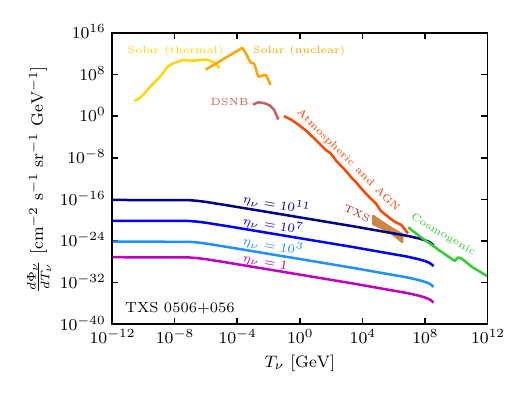}
    \caption{\textit{Upper plot}: All flavour flux of relic neutrinos boosted by Milky Way cosmic ray protons at Earth, for different values of the local neutrino overdensity in the Milky Way. The neutrino mass is taken to be $m_\nu=0.1$ eV. For comparison, we show the expected flux from thermal solar neutrinos \cite{Vitagliano:2017odj}, the measured flux from nuclear solar neutrinos \cite{Bahcall}, diffuse supernova neutrino background \cite{Bisnovatyj, Krauss, Horiuchi:2008jz}, atmospheric \cite{Seckel, Ingelman:1996mj, Super-Kamiokande:2015qek}, and the diffuse extragalactic neutrino background \cite{Murase:2015xka, IceCube:2016umi, IceCube:2020acn}. We also show the current upper limits on the cosmogenic neutrino flux at very high-energies from Auger \cite{PierreAuger:2015ihf}, IceCube \cite{IceCube:2016uab} and ANITA \cite{ANITA:2018vwl}. \textit{Lower plot}:  All flavour flux of
 boosted relic neutrinos from TXS 0506+056 (see main text for details).}
\label{fig:FullSpectrum}
\end{figure}

This work focuses on the phenomenological implications of the boosted relic neutrino flux at very high energies. We find that the boosted relic neutrino flux is maximal at lower neutrino energies, see Fig.~\ref{fig:FullSpectrum}. This prediction on the neutrino flux is unique in the region between the cosmic neutrino background, neutrinos from BBN \cite{Khatri:2010ed, Ivanchik:2018fxy}, and the thermal neutrino flux from the Sun \cite{Haxton:2000xb,Vitagliano:2017odj}. However, the flux at these energies is too small to be discerned, comparable to the diffuse extragalactic neutrino background observed by IceCube.

\section{Conclusions}

We have derived upper limits on the local relic neutrino overdensity and the relic neutrino overdensity in TXS 0506+056, from the consideration that a fraction of the relic neutrinos in the Milky Way and TXS 0506+056 is boosted to high energies due to scatterings with cosmic ray protons. The boosted relic neutrinos can reach the energies of the diffuse extragalactic neutrino background and cosmogenic neutrinos, but their flux is very small. However, if an overdensity of relic neutrinos is present in the Milky Way or in TXS 0506+056, this flux might be detectable due to a differentiated spectral shape w.r.t. the already measured high-energy neutrino fluxes. The non-observation of such excess allows an upper limit to be placed on the relic neutrino density. For a lightest neutrino mass of $m_{\nu} \sim 0.1$ eV, we obtain an upper limit on the local relic neutrino overdensity of $\eta_{\nu} \sim 10^{13}$, which is close to the previously most robust upper limit from KATRIN, and valid over much larger physical scales. For these masses, we also find an upper limit on the neutrino overdensity of TXS 0506+056 of $\eta_{\nu} \sim 10^{10}$, stronger than the bound from KATRIN. We also note that TXS 0506+056 is unlikely to be the only neutrino-emitting quasar. Considering the combined impact of various sources in the Universe may increase the total boosted relic neutrino flux reaching Earth by orders of magnitude. We leave this task for future investigation.

Detecting relic neutrinos is a challenging but crucial task to obtain insights into the cosmological model of the Universe and the neutrino mass generation mechanism. We have discussed the possibility of detecting a boosted component of relic neutrinos by cosmic ray scatterings with high-energy neutrino experiments. We have shown that this detection requires either an improvement in the sensitivity of current neutrino experiments of $\sim 10$ orders of magnitude or an overdensity of relic neutrinos of $\eta_{\nu} \gtrsim 10^{10}$ induced by strong gravitational effects near supermassive black holes or by some BSM mechanism. Although both possibilities remain abject to our view, our result constitutes the strongest bound on the relic neutrino overdensity in a galaxy derived to date and opens a new future avenue for the direct detection of the cosmic neutrino background.

\begin{section}{ACKNOWLEDGEMENTS} 
We are grateful to Chris Cappiello, Juan Manuel Cano, Miguel Escudero, Jérôme Vandecasteele, Carlos Argüelles and Matheus Hostert for useful discussions. MC and GH are supported by the Collaborative Research Center SFB1258 and by the Deutsche Forschungsgemeinschaft (DFG, German Research Foundation) under Germany's Excellence Strategy - EXC-2094 - 390783311. GH is also supported by the the U.S. Department of Energy under the award number DE-SC0020250 and DE-
SC0020262. IMS is supported by the U.S. Department of Energy Office of Science, Office of High Energy Physics, under Award Number DE-SC0020262. 

\end{section}

\bibliography{References}    

\clearpage

\end{document}